\documentclass[twoside,leqno,twocolumn]{article}
\usepackage{amsmath,bm}
\usepackage{array}
\usepackage{ltexpprt}
\usepackage{graphicx}
\usepackage{url}
\usepackage{verbatim}
\usepackage{booktabs}
\usepackage{multirow}
\usepackage{booktabs}
\usepackage{threeparttable}

\begin{document}
\title{\Large Future Influence Ranking of Scientific Literature}
 \author{Senzhang Wang$^\star$
 \and
 Sihong Xie$^\dag$
 \and
 Xiaoming Zhang$^\star$
 \and
 Zhoujun Li$^\star$
 \and
 Philip S. Yu$^\dag$
 \and
 Xinyu Shu$^{\dag\dag}$}
\date{}
 \maketitle
 \begin{abstract}
 \let\thefootnote\relax\footnotetext{\\$^\star$School of Computer Science and Engineering, Beihang University, Beijing, China. \{szwang@cse, lizj@, yolixs@\}.buaa.edu.cn\\
$^\dag$Department of Computer Science, University of Illinois at Chicago, IL, USA. \{sxie6, psyu\}@uic.edu\\
$^{\dag\dag}$International College, China Agricultural University, Beijing, China. shu921@sina.com
 }

\date{} \maketitle
\small\baselineskip=9pt
Researchers or students entering a emerging research area are particularly interested in what newly published papers will be most cited and which young researchers will become influential in the future, so that they can catch the most recent advances and find valuable research directions. However, predicting the future importance of scientific articles and authors is challenging due to the dynamic nature of literature networks and evolving research topics. Different from most previous studies aiming to rank the current importance of literatures and authors, we focus on \emph{ranking the future popularity of new publications and young researchers} by proposing a unified ranking model to combine various available information. Specifically, we first propose to extract two kinds of text features, words and words co-occurrence to characterize innovative papers and authors. Then, instead of using static and un-weighted graphs, we construct time-aware weighted graphs to distinguish the various importance of links established at different time. Finally, by leveraging both the constructed text features and graphs, we propose a mutual reinforcement ranking framework called \emph{MRFRank} to rank the future importance of papers and authors simultaneously. Experimental results on the ArnetMiner dataset show that the proposed approach significantly outperforms the baselines on the metric
\emph{recommendation intensity}.
\end{abstract}
\section{Introduction}

When entering a traditional research area such as data mining or database, researchers may most concern about: ``Which classical papers are most influential and valuable to read in this field?'' and ``Who are the most influential researchers currently?''. But when they entering a newly emerging research direction like social computing or cloud computing, they may be more interested in: ``What latest papers may become popular in the future so that I should read?'' and ``Which young researchers will probably become influential so that I should follow their works?''. While traditional ranking models mainly focus on addressing the former two problems, this paper aims to tackle the latter two. However, it is extremely hard for new comers to a newly emerging research area to find the latest valuable publications and identify potentially influential researchers. It is due in part to the tremendous number of papers published each year, making searching for excellent new papers time-consuming, and in part due to the dynamic nature of the latest research topics and evolving literature networks.

Previous works on scientific literature ranking can be mainly divided into citation count based methods [7-10] and graph-based ranking methods [1], [2], [4], [11-15]. Citation count is a simple but useful measurement to rank the importance of papers and authors [7], [8]. Then some more complicated metrics are proposed, such as h-index [9]. But such methods ignore the available structure information such as citation and coauthor graphs, which is also important to measure the influence of papers and authors. Recently, many studies have focused on applying graph-based methods to literature ranking [1], [2], [14]. For example, Zhou et, al. proposed to combine citation, authorship and co-authorship networks to simultaneously rank publications and authors [1]. Jiang et, al. leveraged networks of papers, authors, and venues to set up a unified mutual reinforcement model to rank papers, authors and venues [2]. Graph-based methods can usually obtain more reasonable ranking results, because they take both the popularity (citation count) and prestige (link information) of publications into consideration.

However, there are two challenges to address the proposed issue. The first one is how to model the dynamic and evolving literature networks such as citation or co-authors networks. Most previous works ignore the dynamic property of graphs, such that the ranking result is usually biased to old articles. The top ranked papers are usually overwhelmed by the classical ones published many years ago. Similar problem also exists in authors ranking. Therefore, existing methods are not suitable to rank the valuable new papers and influential young researchers. Although some efforts have been made to explore additional information, such as time information [13], [15], they are not effective to model the evolving citation or coauthor links. To address this challenge, we propose the time-aware weighted networks to profile the dynamic nature of various literature graphs. We believe that newly established links are more useful to predict their future trend. For example, we discover that the papers which are frequently cited by new papers may probably continue to obtain many citations in the near future than those whose most citations are old.

The second challenge is how to use the paper content information to help identify potentially influential new papers. Previous works ignore the content information. But the content information is also important for measuring the quality of papers, especially for the new ones with only a handful of citations currently. Generally, more innovative papers are more likely to address new problems or discuss new topics. Thus such papers contain more novel text features. For example, social media related texts such as ``social network'', ``social media'', ``twitter'' and ``Facebook'' are quite popular in recent publications. Early papers on this topic are very innovative, and most of them have obtained hundreds of citations. Therefore, effectively identifying the pioneering papers by capturing their novel text features may largely help us find potentially influential papers and corresponding researchers. In this paper, we first present an burst detection based method to measure the innovative degree of two kinds of text features, words and words co-occurrence. Then, by mapping the text features and papers or authors to bipartite graphs, we construct the paper-text features and author-text features graphs.

By combining the above constructed graphs, we finally propose a unified ranking model \emph{MRFRank}. \emph{MRFRank} is a HITS like algorithm which employs the mutual reinforcement relationships across networks of papers, authors and text features. The intuition is that future influential researchers using many novel text features of rising popularity lead to the future important papers; and future important papers containing many novel text features of rising popularity lead to future influential researchers. The main contributions of the paper can be summarized as follows. 

\begin{itemize}
\item[$\bullet$]
{We propose to characterize the innovative papers and authors by their innovative text features. To find innovative text features, we use a burst detection based method to measure their innovative degree. To the best of our knowledge,
this is the first attempt to incorporate the paper content information into literature ranking.}
\item[$\bullet$]
{To capture the dynamic and evolving nature, we use the time information in both citation and coauthor graphs. The proposed ranking algorithm is conducted on the time-aware weighted networks instead of the original static graphs.}
\item[$\bullet$]
{A unified ranking model named MRFRank is proposed by incorporating the extracted text features and constructed weighted graphs. As a mutual reinforcement ranking framework, MRFRank ranks the future influence of papers, authors and text features simultaneously.}
\item[$\bullet$]
{We empirically evaluate our approach on the ArnetMiner dataset. The results demonstrate that our method outperforms existing state-of-the-art algorithms, including FutureRank and MutualRank on ranking new papers and young researchers.}
\end{itemize}

The rest of this paper is organized as follows. Next we will review related works. Section 3 will describe how we model the time and content information. Then, we will introduce the unified ranking model in section 4. The experiment and evaluation are given in section 5. Finally, we conclude this paper in section 6.

\section{Related Work}
The earliest work on scientific literature ranking was the citation count method proposed by Garfield [7]. Though very simple, citation count is widely used to measure the importance of papers and researchers. Based on citation count, several more complicated metrics are proposed, such as h-index proposed by Hirsh [9] and its variation g-index proposed by Egghe [10].

The main limitation of above methods is that they only consider articles' popularity but ignore their prestige.  With the popularity of PageRank algorithm, many studies tried to apply PageRank approach to the literature network to rank papers or authors. For example, Ding et al. proposed to apply the PageRank algorithm on the co-author network to rank the influence of researchers [12]. Bollen et al. applied the PageRank method to the citation network to rank the importance of articles [15]. 

PageRank method can only work on one type of network, which limits its effectiveness in ranking different kinds of objects. Recent works began to consider exploring heterogeneous networks to rank multiple entities simultaneously [1], [2], [14]. For example, the Co-Rank algorithm proposed by Zhou and Orshanskiy combined the citation network and co-authorship network to improve the ranking results for both authors and articles [1]. Similarly, Jiang et al. proposed a unified mutual reinforcement ranking model which involves intra- and inter- network information for ranking papers, authors and venues [2]. These methods benefit from different graphs, and therefore can usually achieve better ranking results.

Some efforts have also been made to rank the future popularity of publications [4], [6], [11]. Walker et al. proposed to add the publication time of the articles to the ranking model to predict the future citation count of papers [11]. Similarly, FutureRank aims to predict the future popularity of scientific articles [4]. Another most recent related work is conducted by Wang et al. [6]. They add the time information to the author-paper relationship to rank the future citations of papers. However, the limitation of above works is that the time information is not fully utilized in various literature networks. For example, the citation and coauthor relationships are also time sensitive, but no work has studies these properties, to the best of our knowledge.

\section{Modeling the Content and Time Information}
In this section, we will introduce how to model the time and content information to help us better rank the future influence of scientific literature. First, we will explain why the content information is helpful. Then we will propose to use two types of text features, words and words co-occurrence to characterize papers and authors. For each text feature, we propose a burst detection based method to quantitatively measure its innovative degree. Next, we will present how to construct the time-aware citation and co-author relationships. Before formulating our approach, we first give some notations used in the paper in Table 1.
\newcommand{\tabincell}[2]{\begin{tabular}{@{}#1@{}}#2\end{tabular}}
\begin{table}[htb!]\small
\centering
\caption{Notations}
\begin{tabular}{|l|l|}
\hline
Notation & Description\\
\hline
$\bm{P}$ & The set of paper collection.\\
\hline
$\bm{A}$ & The set of author collection.\\
\hline
$\bm{F}$ & The set of text feature collection.\\
\hline
$\bm{E}$ & \tabincell{l}{The vector indicating the innovative degree \\of text features.}\\
\hline
$\bm{A\_P}$ & \tabincell{l}{The vector indicating the future \\authority of papers.}\\
\hline
$\bm{A\_A}$ & \tabincell{l}{The vector indicating the future \\authority of authors.}\\
\hline
$\bm{A\_F}$ & \tabincell{l}{The vector indicating the future \\authority of text features.}\\
\hline
$\bm{M^{PP}}$ & \tabincell{l}{The $|N|\times|N|$ matrix indicating \\citation graph.}\\
\hline
$\bm{M^{AA}}$ & \tabincell{l}{The $|M|\times|M|$ matrix indicating \\coauthor graph.}\\
\hline
$\bm{M^{PF}}$ & \tabincell{l}{The $|N|\times|K|$ matrix indicating \\paper-text feature graph.}\\
\hline
$\bm{M^{AF}}$ & \tabincell{l}{The $|M|\times|K|$ matrix indicating \\author-text feature graph.}\\
\hline
\end{tabular}
\end{table}
\subsection{Text feature extraction}
When a new research topic emerges, only a small number of researchers focus on it and publish related papers. Then gradually, more and more researchers become interested in it and begin to follow the pioneers' works. Finally, with more and more related papers published, the topic is not fresh any more; and many researchers turn to other new problems. From the perspective of citation count, papers published early are more likely to get numerous citations, because more and more papers published later cite them. Contrarily, it becomes harder and harder for the latecomers to get citations, since the topic is outdated and too many related papers already.

As an example, Fig. 1 shows the number of yearly published papers whose titles contain ``associate rule'' and their corresponding average citation counts from 1994 to 2010. The left figure demonstrates that the paper number increases rapidly in the first decade, and reaches its peak in 2008. Then, it begins to decrease. Interestingly, the trend pattern shown in the right figure is almost the opposite to the left. The most cited papers are those published very early. In fact, the two figures are not surprising and consistent with our intuition. More innovative papers are usually the earlier works addressing new research issues. Hence, such papers are much easier to obtain high number of citations.
\begin{figure}[!tp]
\centering
\setlength{\belowcaptionskip}{-0.5cm}
\includegraphics[height=2.8cm]{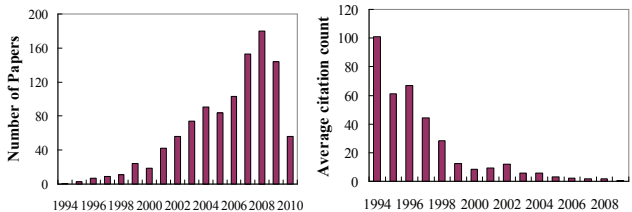}
\caption{yearly published papers \emph{vs} citation count.}
\end{figure}

Therefore, effectively identifying the early papers about emerging new topics may greatly help us to recognize their potential popularity and guide us on selecting research directions. But the challenge is: how to find the pioneering papers early on? Finding a few groundbreaking papers from a large volume of collection is indeed a non-trivial task, but we can find some more innovative ones based on their text features. Next we will describe what text features we will extract and how to measure their innovativeness.

First, two types of text features, words and words co-occurrence in the same sentence are extracted from the titles and abstracts of papers. Some new words emerge and become hot with the emerging of new topics. For example, social networks have gained significant popularity and many related papers are published each year. Most of these papers may contain the following words: "twitter", "Facebook", "Weibo" and "social media". Most of these words are new relative to traditional topics. We also use words co-occurrence as the text feature. Two explosive co-occurring words may imply the combination of two different topics that may be very innovative. For example, the word pairs ``deep-sentiment'' and ``learning-sentiment'' may imply the combination of the topics ``deep learning'' and ``sentiment analysis''.

Then, we propose to measure the innovativeness of text features by a burst detection based method. Burst detection is widely used in event detection in social media [3], [5], [6]. Here we apply this technique to measure the innovative degree of each text feature. In event detection, a term is defined as bursty if it frequently occurs in a specified time window but rarely occurs in the past [6]. Similarly, we say the paper's text feature is innovative if its frequency increases remarkably in a specified time window. 

We assume the frequency of text features follows the Poisson distribution
\begin{equation}
f^i(k,\lambda)=P(x_i=k)=\frac{\lambda_i^ke^{-\lambda_i}}{k!}
\end{equation}
where $i$ denotes the $i_{th}$ text feature, $x_i$ is the frequency of text feature $i$, and $\lambda_i$ is the mean of the variable $x_i$. The maximum likelihood estimation of $\lambda_i$ is the sample mean $\widetilde{\lambda_i}=\frac{1}{n}\sum_{i=1}^{n}x_i$, and the maximum likelihood estimation of all the features is $\widetilde\lambda=\frac{1}{K}\sum_{i=1}^{K}\widetilde\lambda_i$.
Given the feature frequency $x_{i}^{<t_{j-1},t_{j}>}$ in the $j_{th}$ time window $<t_{j-1},t_j>$, we define the degree of innovativeness of feature $x_i$ as
\begin{equation}
\begin{split}
E_{i}^{<t_{j-1},t_j>}=&\frac{|x_{i}^{<t_{j-1},t_j>}-\widetilde{\lambda_i}|}{\widetilde{\lambda}}\cdot\\
&[\sum_{s=1}^{u}(\frac{x_{i}^{<t_{j-1},t_j>}-x_{i}^{<t_{j-s-1},t_{j-s}>}}{\widetilde{\lambda_i}}) \frac{1}{s}]\cdot\\
&e^{-\rho(t_j-t_0)}.
\end{split}
\end{equation}
This measurement contains three parts. The first part is the absolute value between feature frequency $x_{i}^{<t_{j-1},t_j>}$ and the estimated mean frequency $\widetilde{\lambda_i}$. It means that a higher feature frequency will benefit its innovativeness. The second part is the difference between the feature's current frequency $x_{i}^{<t_{j-1},t_j>}$ and the frequencies in its nearest past $s$ time windows. $u$ is a parameter which limits the number of previous time windows and is set to 3. This part means if the current feature frequency has a significant increment compared with its previous $u$ nearest neighbors, its innovative degree is considered to be high. Meanwhile, to highlight the very early features occur recently, we use a time-weighted exponential function as the third part. $t_0$ is the time when the text feature first appears in the papers collection. Fig. 2 shows an illustration of our idea.
\begin{figure}[!tp]
\centering
\setlength{\belowcaptionskip}{-0.5cm}
\includegraphics[height=5cm]{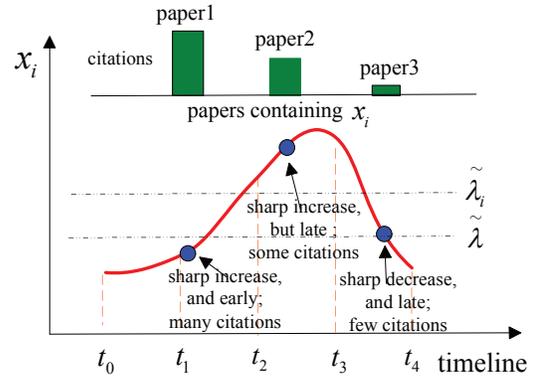}
\caption{An illustration of innovativeness measure.}
\end{figure}

Based on the extracted text features and their innovative degrees, the papers and authors can be characterized as follows.

\textbf{Paper.} The paper $P_i$ can be characterized as a set of text features $\{f_i^1,f_i^2,...,f_i^n\}$. Each text feature can be denoted as a triple $(w_i,E_i^{<t_{j-1},t_j>},t_i)$, where $w_i$ is the $tf-idf$ weight of the feature $i$, $E_i^{<t_{j-1},t_j>}$ is the innovative degree of feature $i$ in the $j_{th}$ time window $<t_{j-1},t_j>$, and $t_j$ is the paper published time.

\textbf{Author.} The author $A_i$ can also be characterized as a set of text features $\{f_i^1,f_i^2,...,f_i^m\}$. Each feature here can be denoted as such a tuple $(w_i,E_i^{<t_{j-1},t_j>})$. Here $w_i$ is $tf-idf$ like weight of feature $i$, and $E_i^{<t_{j-1},t_j>}$ is the innovative degree of feature $i$ in the $j_{th}$ time window.

\subsection{Time-aware Citation and Coauthor Relationships} Several prior studies have tried to use the time information for literatures ranking [4], [13], [15]. For example, to predict the future prestige of papers, Hassan et, al. assume that newly published papers are more likely to get more citations than older ones in the future [4]. Nevertheless, not all the new papers will obtain more citations than old ones. Actually, most papers, no matter new or old, get a small number of citations. Only a few papers are frequently cited. Therefore, instead of utilizing the publishing time, we apply the time when links established, such as the time when a paper cites another paper. 
\begin{figure}[!tp]
\centering
\setlength{\belowcaptionskip}{-0.5cm}
\includegraphics[height=4.5cm]{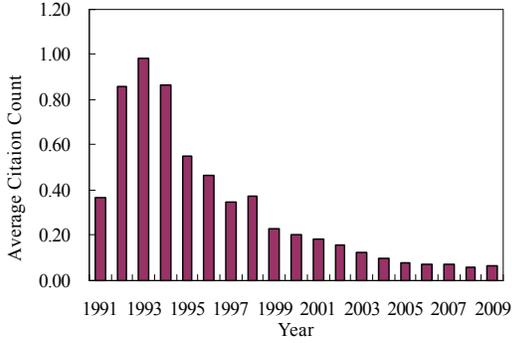}
\caption{Average \# of citations in each year.}
\end{figure}

We assume that the papers frequently cited recently are much more likely to keep obtaining new citations than those whose citations are mostly old. Fig. 3 shows the statistics of the average citation counts in each year of all the papers published in 1990. We can see that the citation counts curve is rather smooth, and papers are statistically unlikely to obtain many citations in the future if they are not attractive in recent years. To take this into account, we propose the time-aware weighted citation relationship to highlight new links.

We denote the paper $P_i$ citing paper $P_j$ at time $T_{cite}$ as $C_{i\rightarrow j}(T_{cite})$ with value 1. We propose to utilize the following exponentially decaying equation to measure the weight of citation relationship between $P_i$ and $P_j$.
\begin{equation}
TW_{P_i\rightarrow P_j}=e^{-\rho (T_{current}-T_{cite})} C_{i\rightarrow j}(T_{cite}).
\end{equation}
Here $\rho$ is a predefined decaying parameter.

Similarly, a researcher who recently co-authors with influential is more likely to keep co-authoring new papers with them. We denote the author $A_i$ co-authors the paper $P_k$ with $A_j$ at time $T_{co}$ as $A_{i-j}^{P_k}(T_{co})$ with value 1. The time-aware weighted co-author relationship between them on $P_k$ can be represented as
\begin{equation}
TW_{A_i-A_j}^{P_k}=e^{-\rho (T_{current}-T_{co})} A^{P_k}_{i-j}(T_{co}).
\end{equation}
Additionally, two authors may co-author many papers, thus the time weighted co-author relationship between them over all the papers can be denoted as
\begin{equation}
\begin{split}
TW_{A_i-A_j}=\sum_{P_k\in co(A_i,A_j)}e^{-\rho (T_{current}-T_{co}^{P_k})} A^{P_k}_{i-j}(T_{co})
\end{split}
\end{equation}
where $co(A_i,A_j)$ is the set of papers coauthored by $A_i$ and $A_j$.
\section{MRFRank: The Unified Ranking Model} In this section, we introduce how to integrate the time-aware weighted graphs and rich texts into a unified \underline{M}utual \underline{R}einforcement model for ranking the \underline{F}uture importance of scientific articles and authors simultaneously (MRFRank). Our model is based on the following dependency rules:
\begin{itemize}
\item[$\bullet$]
{Influential papers are frequently cited by other influential papers; are usually written by well-know researchers; and contain many innovative text features.}
\item[$\bullet$]
{Influential researchers publish many high quality papers; often co-author papers with other influential researchers; and always catch up the most recent advances, thus their papers usually contain many innovative text features.}
\item[$\bullet$]
{Recent citations are more indicative of papers' future citations; and the influence of recent co-authors is more indicative of the influence of authors' future co-authors.} 
\end{itemize} 
\subsection{Literature Graphs Construction} Before describing the algorithm in detail, we first give a brief introduction to the graphs used in our approach. There are three types of nodes, i.e. authors, papers, and text features, forming five types of graphs, i.e. coauthor graph, paper citation graph, author-paper graph, author-text feature graph, and paper-text feature graph.

\textbf{Time-aware coauthor Graph.} There exists an edge $e_{ij}$ if $A_i$ and $A_j$ coauthor at least one paper. The matrix representation can be defined as
\begin{equation*}
M^{AA}_{ij} = \left\{
             \begin{array}{lcl}
             {TW_{A_i-A_j}} &\text{if $A_i$ coauthors papers with $A_j$}\\
             {0} &\text{otherwise}
             \end{array}
        \right.
\end {equation*}
The coauthor network is a time-aware weighted graph, and the weight of each edge is defined in (3.5). 

\textbf{Time-aware paper citation graph.} 
There exists an edge $e_{ij}$ if paper $P_i$ cites paper $P_j$. The adjacency matrix of the graph is denoted as
\begin{equation*}
M^{PP}_{ij} = \left\{
             \begin{array}{lcl}
             {TW_{P_i\rightarrow P_j}} &\text{if paper $P_i$ cites $P_j$}\\
             {0} &\text{otherwise}
             \end{array}
        \right.
\end {equation*}
The citation graph is also a time-aware weighted graph. The weight of each edge is defined in (3.3). 

\textbf{Author-paper graph.} This graph contains two kinds of nodes, papers and authors. If $A_i$ is the author of paper $P_j$, there exists an edge $e_{ij}$. It is a bipartite graph, and the matrix representation can be denoted as
\begin{equation*}
M^{AP}_{ij} = \left\{
             \begin{array}{lcl}
             {1} &\text{if $A_i$ is the author of paper $P_j$}\\
             {0} &\text{otherwise}
             \end{array}
        \right.
\end {equation*}

\textbf{Paper-text feature graph.} This graph also contains two kinds of nodes, papers and text features. 
If paper $P_i$ contains the text feature $f_j$, there exists an edge. Its matrix representation is
\begin{equation*}
M^{PT}_{ij} = \left\{
             \begin{array}{lcl}
             {w_{ij}} &\text{if paper $P_i$ contains the feature $f_j$}\\
             {0} &\text{otherwise}
             \end{array}
        \right.
\end {equation*}
$w_{ij}$ is the $tf-idf$ weight of text feature $f_j$ in paper $P_i$.

\textbf{Author-text feature graph.} This graph contains authors and text features. If author $A_i$ uses the text feature $f_j$ as least once in his papers, there exists an edge between them. The matrix representation is \begin{equation*}
M^{AT}_{ij} = \left\{
             \begin{array}{lcl}
             {w_{ij}} &\text{if $A_i$ uses the feature $f_j$}\\
             {0} &\text{otherwise}
             \end{array}
        \right.
\end {equation*}
$w_{ij}$ is the $tf-idf$ like weight of $f_j$ used by $A_i$.

\subsection{Algorithm}
We conduct the MRFRank algorithm iteratively on the above graphs by the following steps:

1) Initially, the authority vectors of $\bm{A\_P}$, $\bm{A\_A}$, and $\bm{A\_F}$ are set to $\frac{\bm{I}_N}{N}$, $\frac{\bm{I}_M}{N}$, and $\frac{\bm{I}_K}{K}$. $\bm{I}_N$, $\bm{I}_M$, and $\bm{I}_K$ are unit vectors. 
Then repeat steps 2)-4) until it converges.

2) Based on the dependency rules, update the paper authority vector $\bm{A\_P}^{t+1}$ by the authority vectors of authors $\bm{A\_A}^t$, papers $\bm{A\_P}^t$ and text feature $\bm{A\_F}^t$; the author-paper matrix $\bm{M^{AP}}$, the paper citation matrix $\bm{M^{PP}}$, and the paper-text feature matrix $\bm{M^{PT}}$.

3) Update the author authority vector $\bm{A\_A}^{t+1}$ by the authority vectors of papers $\bm{A\_P}^t$, authors $\bm{A\_A}^t$, and text features $\bm{A\_F}^t$; the coauthor matrix $\bm{M^{AA}}$, the author-paper matrix $\bm{M^{AP}}$, and the author-text feature matrix $\bm{M^{AT}}$.

4) Update the text feature authority vector $\bm{A\_F}^{t+1}$ by the the authority vectors of papers $\bm{A\_P}^t$ and authors $\bm{A\_A}^t$; the innovative degree $\bm{E}$ of features, the paper-text features matrix $\bm{M^{PT}}$, and the author-text features matrix $\bm{M^{AT}}$.

Specifically, the iteration process of the MRFRank algorithm can be formulated as follows.\\
\textbf{future authority of paper $P_i$}
\begin{equation}
\begin{split}
&A\_P_{i}^{t+1}=\alpha_p \sum_{P_j\in Cite(P_i)}M_{ij}^{PP} A\_P_{j}^t+\\
&\beta_p (1-\alpha_p)\sum_{A_j\in Author(P_i)}M_{ij}^{PA} A\_A_{i}^t+\\
&(1-\beta_p)(1-\alpha_p) \sum_{f_j\in Feature(P_i)}A\_F^t_{j} M_{ij}^{PT}
\end{split}
\end{equation}
\textbf{future authority of author $A_i$}
\begin{equation}
\begin{split}
&A\_A_{i}^{t+1}=\alpha_a \sum_{A_j\in Coauthor(A_i)}M_{ij}^{AA} A\_A_{j}^t+\\
&\beta_a (1-\alpha_a)\sum_{A_i\in Author(P_j)}M_{ij}^{AP} A\_P_{j}^t+\\
&(1-\beta_a)(1-\alpha_a) \sum_{f_j\in Feature(A_i)}A\_F^t_{j} M_{ij}^{PT}
\end{split}
\end{equation}
\textbf{future authority of text features $f_i$}
\begin{equation}
\begin{split}
&A\_F_{i}^{t+1}=[\alpha_f \sum_{A_j\in Author(f_i)}M_{ij}^{TA} A\_A_{j}^t+\\
&(1-\alpha_f)\sum_{P_j\in Paper(f_i)}M_{ij}^{TP} A\_P_{j}^t] E_k
\end{split}
\end{equation}
Here $Cite(P_i)$ denotes the set of papers citing $P_i$. $Author(P_i)$ denotes the set of authors of $P_i$. $Feature(P_i)$ denotes the set of text features in $P_i$. $Coauthor(A_i)$ denotes the set of $A_i$'s coauthors. $Feature(A_i)$ denotes the set of text features used by $A_i$. $Author(f_i)$ denotes the set of users who use the text feature $f_i$. $Paper(f_i)$ denotes the set of papers containing text feature $f_i$.

(4.6), (4.7) and (4.8) can be rewritten in matrix forms as follows,
\begin{equation}
\begin{split}
\bm{A\_P}^{t+1}=&\alpha_p (\bm{M^{PP}} \bm{A\_P}^t)+\\
&\beta_p (1-\alpha_p) (\bm{M^{PA}} \bm{A\_A}^t)+\\
&(1-\beta_p)(1-\alpha_p) (\bm{M^{PT}} \bm{A\_F}^t)
\end{split}
\end{equation}
\begin{equation}
\begin{split}
\bm{A\_A}^{t+1}=&\alpha_a (\bm{M^{AA}} \bm{A\_A}^t)+\\
&\beta_a (1-\alpha_a) (\bm{M^{AP}} \bm{A\_P}^t)+\\
&(1-\beta_a)(1-\alpha_a) (\bm{M^{AT}} \bm{A\_F}^t)
\end{split}
\end{equation}
\begin{equation}
\begin{split}
\bm{A\_F}^{t+1}=&[\alpha_f (\bm{M^{TA}} \bm{A\_A}^t)+\\
&(1-\alpha_f)(\bm{M^{TP}} \bm{A\_P}^t)] \bm{E}
\end{split}
\end{equation}
After every iteration, we normalize each vector by dividing the sum of all its elements. (4.9), (4,10) and (4,11) can be further rephrased as the following equation
\begin{equation}
\bm{R}^{t+1}=\bm{M}\bm{R}^t
\end{equation}
where $\bm{R=[A\_P^T, A\_A^T, A\_F^T]^T}$, and
\begin{equation}
\bm{M}=
\end{equation}
$$\begin{pmatrix}
\begin{smallmatrix}
\alpha_p\bm{M^{PP}}\bm{\Lambda_I} & \beta_p (1-\alpha_p) \bm{M^{PA}}  & (1-\beta_p) (1-\alpha_p) \bm{M^{PT}} \\
 \beta_a (1-\alpha_a) \bm{M^{AP}} & \alpha_a \bm{M^{AA}} \bm{\Lambda_I} & (1-\beta_a) (1-\alpha_a) \bm{M^{AT}} \\
 (1-\alpha_f) \bm{\Lambda_E} \bm{M^{TP}} & \alpha_f \bm{\Lambda_E} \bm{M^{TA}} & \bm{\Lambda_0}
\end{smallmatrix}
\end{pmatrix}$$
$\bm{\Lambda_I}$ and $\bm{\Lambda_E}$ are both diagonal matrixes with the diagonal elements $\Lambda_{ii}=1$ and $\Lambda_{ii}=E_i$, respectively. $\bm{\Lambda_0}$ is a zero matrix. The matrix $\bm{M}$ is a transition matrix corresponding to a Markovian process, thus it is not hard to verify that $\bm{R}$ is the eigenvector of matrix $\bm{M}$, and it will converge to the primary eigenvector. 

\section{Experiment and Evaluation}
\subsection{Dataset}
The publicly available ArnetMiner dataset on paper publications\footnote{\url{http://arnetminer.org/citation#b541}} is used to evaluate our approach. It contains 1,572,277 papers published before 2011 and 2,084,019 corresponding citations. The metadata of each paper contain paper ID, title, abstract, authors, publication year, publication venue, cited papers ID and citation count.

We preprocess the dataset as follows. First, as we only rank research papers, the survey papers are eliminated. Second, the papers with no citations and do not cite other papers are removed. Third, the collection contains some workshop proceedings. These proceedings contain all the papers published in the workshop. But in the dataset, the whole proceeding is considered as a ``paper''. Such proceedings are also removed. In addition, the metadata of most old paper are incomplete. For example, most papers published before 1990 have no citation relationships and abstracts. Therefore, we remove the papers published before 1990 and the papers published after 1990 but with incomplete metadata. After the preprocessing, there are 302,336 remained papers and 1,085,181 remained citations.
\subsection{Ground Truth and Baselines}
As we aim to rank the future influence of papers and authors, we adopt the number of the their future citations as the ground truth [15]. Then the ground truth ranking of papers and authors can be obtained by sorting them in the descending order of their future citation counts. We first divide the dataset into two parts: ranking part and evaluation part. Specifically, we select and rank the papers published before 2005 to obtain the ranking lists of papers and authors, then the ground truth ranks are obtained by ranking their citation counts from 2005 to 2011. Meanwhile, in order to find what new papers will be the most cited, we select the papers published in the same recent year from the entire ranking list for evaluation. For example, for all the papers published in 2000, which ones will become the most cited? Similarly, we define young researchers as those who begin to publish papers from a specific recent year. For example, for all the researchers starting to publish papers from 2000, who will become influential? Thus we only select those starting to publish paper in the same year from the entire ranking list for evaluation. 

We use the \textbf{\emph{recommendation intensity (RI)}} proposed by Jiang et, al. as the evaluation metric [2]. Assume $R$ is the list of top-$k$ returned papers/authors of a ranking approach, and $L$ is the list of ground truth, then for each paper/author $P_i$ in $R$ with the ranked order $o_r$, the \emph{recommendation intensity} of $P_i$ at $k$ can be defined as
\begin{equation}
RI(P_i)@k=\left\{
             \begin{array}{lcl}
             {1+(k-o_r)/k} &P_i\in L\\
             {0} &P_i\notin L
             \end{array}
        \right.
\end {equation}
It means that if the paper/author $P_i$ is in the top-$k$ ground truth list $R$ and is ranked higher (smaller $o_r$), then its \emph{recommendation intensity} is higher.

Based on each paper's/author's \emph{recommendation intensity} in the list $R$, the \emph{recommendation intensity} of the list $R$ at $k$ can be defined as
\begin{equation}
RI(R)@k=\sum_{P_i\in R}RI(P_i)@k.
\end {equation}

\textbf{FutureRank (FR)} [4] and \textbf{MutualRank (MR)} [2] are selected as baselines. FutureRank is a representative method to predict the future important papers proposed recently. MutualRank is the state-of-art graph-based method that integrates mutual reinforcement relationships among several graphs to rank papers, authors, and venues simultaneously. Additionally, in order to study how much performance can be improved by using the time or content information, we use the following three variations of the proposed approach as baselines: MRFRank without time information \textbf{(MRFR-T)}, MRFRank without content information \textbf{(MRFR-C)}, and MRFRank without time and content information \textbf{(MRFR-TC)}.

\subsection{Case study}
We first give a case study in Tables 2 and 3. Table 2 lists the titles of top-10 papers returned by MRFRank in 2000, and their corresponding rankings in the ground truth of 2000. We also list the rankings in the ground truth of the top-10 papers returned by FutureRank and MutualRank. It shows the result returned by MRFRank is significantly better than that returned by baselines. 6 out of the top-10 papers returned by MRFRank are in the top-10 rankings of the ground truth, compared with 4 by FutureRank and only 3 by MutualRank. Our approach identifies the papers \emph{``Mining Frequent Patterns without Candidate Generation''} and \emph{``Privacy-Preserving Data Mining''}, which turn out to be very influential, while the two baselines fail to give them high rankings. This is because in 2000, the topics the two papers discussed are very new, and our approach captures their novel text features. Table 3 lists the names of top-10 authors returned by MRFRank and their rankings in the ground truth of 2000. We also list the rankings in the ground truth of the top-10 authors identified by MutualRank and citation count. Our approach also performs best. 8 out of the top-10 authors returned by our approach are in the top-10 rankings of the ground truth. Citation count identifies 6 and MutualRank identifies only 4.

\begin{table*}[htb!]\small
\caption{Top-10 Papers (published in 2000)}
\centering
\setlength{\abovecaptionskip}{5pt}
\begin{threeparttable}
\begin{tabular}{|l|l|l|l|l|}
\hline
\multirow{2}{*}{Titles of top-10 papers returned by MRFRank} &\multirow{2}{*}{Publication Venue} & \multicolumn{3}{l|}{Rankings in GT\tnote{1}}\\
\cline{3-5}
& & MRFR & MR & FR\\
\hline

1: The Gaia Methodology for Agent-Oriented Analysis and Design & AAMAS & 16 & 6 & 14\\
\hline
2: The Cricket location-support system & MobiCom & 4 & 7 & 1\\
 \hline
3: Content-Based Image Retrieval at the End of the Early Years & \tabincell{l}{IEEE Trans. Pattern\\Anal. Mach. Intell.} & 2 & 1 & 2\\
 \hline
4: Text Classification from Labeled and Unlabeled Documents using EM & Machining Learning & 10 & 17 & 10\\
 \hline
\tabincell{l}{5: Energy-Efficient Communication Protocol for\\Wireless Microsensor Networks} & HICSS & 1 & 28 & 4\\
 \hline
6: Equation-based congestion control for unicast applications & SIGCOMM & 18 & 23 & 16\\
 \hline
7: Mining Frequent Patterns without Candidate Generation & SIGMOD & 3 & 16 & 18\\
 \hline
8: Computing with Membranes & J. Comput. Syst. Sci. & 12 & 25 & 126\\
 \hline
9: Privacy-Preserving Data Mining & SIGMOD & 5 & 13 & 11\\
 \hline
10: A scalable location service for geographic ad hoc routing & MobiCom & 14 & 12 & 21\\
\hline
\end{tabular}
\begin{tablenotes}
\footnotesize
\item[1] GT, MRFR, MR, and FR stand for ground truth, MRFRank, MutualRank, and FutureRank, respectively.
\end{tablenotes}
\end{threeparttable}
\end{table*}
\vspace*{0pt}
\begin{table}[htb!]\small
\caption{Top-10 Researchers (start from 2000)}
\centering
\setlength{\belowcaptionskip}{-10pt}
\begin{threeparttable}
\begin{tabular}{|l|l|l|l|}
\hline
\multirow{2}{*}{Top-10 authors by MRFR} & \multicolumn{3}{l|}{Rank order in GT}\\
\cline{2-4}
 & MRFR & MR & CC\tnote{1}\\
\hline

1: Chalermek Intanagonwiwat & 5 & 8 & 5\\
\hline
2: Jian Pei & 2 & 2 & 1\\
\hline
3: Szymon Rusinkiewicz & 10 & 1 & 4\\
\hline
4: Adrian Perrig & 1 & 19 & 3\\
\hline
5: Brad Karp & 4 & 37 & 14\\
\hline
6: Robert Szewczyk & 3 & 6 & 10\\
\hline
7: Mayur Datar & 9 & 36 & 13\\
\hline
8: Wendi Rabiner Heinzelman & 6 & 25 & 42\\
\hline
9: Ya Xu & 55 & 39 & 9\\
\hline
10: Paramvir Bahl & 13 & 24 & 17 \\
\bottomrule
\end{tabular}
\begin{tablenotes}
\footnotesize
\item[1] CC stands for the citation count method.
\end{tablenotes}
\end{threeparttable}
\end{table}

\subsection{Quantitative Comparison} Next, we quantitatively compare the performance of the proposed approach with baselines. Fig. 4 shows the result of the papers published in 2000 and 2001. Fig. 5 shows the result of the researchers who start to publish papers in 2000 and 2001. The figures show that for both researcher and paper rankings, the proposed approach outperforms baselines over various $k$. For the ranking of papers, FutureRank is generally better than MutualRank, but inferior to the proposed approach. MRFRank outperforms FutureRank by at most 10\% on the ranking of papers published in 2000 and 8\% on the ranking of papers published in 2001. For authors ranking shown in Fig. 5, MutualRank is surprisingly no better than simply counting current citations. Our approach outperforms the baselines by at most 20\% in 2000 and 30\% in 2001 for authors ranking.
\begin{figure}[!tp]
\centering
\setlength{\belowcaptionskip}{-0.2cm}
\setlength{\abovecaptionskip}{5pt}
\includegraphics[height=2.7cm]{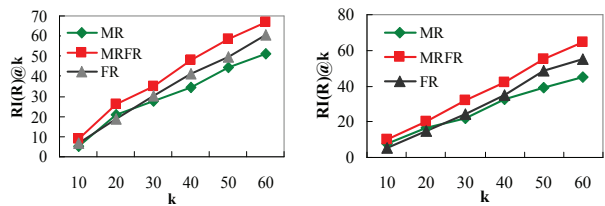}
\caption{RI(R)@k of ranked papers published in 2000 (left) and 2001 (right)}
\end{figure}
\begin{figure}[!tp]
\centering
\setlength{\belowcaptionskip}{-0.5cm}
\setlength{\abovecaptionskip}{5pt} 
\includegraphics[height=2.85cm]{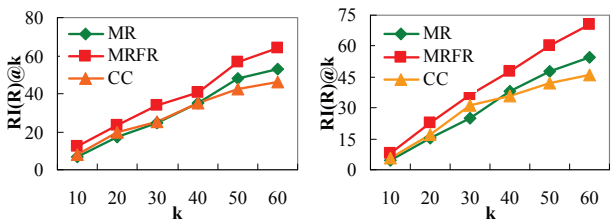}
\caption{RI(R)@k of ranked authors starting publishing papers from 2000 (left) and 2001 (right)}
\end{figure}    

\textbf{Comparison with three variations of MRFRank.} To investigate whether and to what extent the time and content information can improve the performance. We conduct experiments to compare MRFRank with MRFR-T, MRFR-C and MRFR-TC. The result is given in Table 4. The boldface figures are the best results. We can see that in most cases the time and content information do help us get better rankings. It also shows that the results of 2000 and 2001 are much better that that of 2002 and 2003. This is mainly because we only use the available data before 2005 for ranking. Papers have not obtained sufficient citations, and authors have not published many papers in such a short time.

\vspace*{0pt}
\begin{table}[htb!]\small
\caption{Experiment Results}
\centering
\begin{tabular}{|p{0.5cm}|p{1.6cm}|p{0.45cm}|p{0.45cm}|p{0.45cm}|p{0.45cm}|p{0.45cm}|p{0.45cm}|}
\hline
\multirow{2}{*}{year} &\multirow{2}{*}{method} & \multicolumn{2}{l|}{k=10}  & \multicolumn{2}{l|}{k=20} & \multicolumn{2}{l|}{k=50}\\
\cline{3-8}
& & P & A & P & A & P & A\\
\hline
\multirow{4}{*}{2000} & MRFR & \textbf{9} & \textbf{11.9} & \textbf{26} & \textbf{23.4} & \textbf{58.7} & \textbf{56.5}\\
\cline{2-8}
& MRFR-T & 7.4 & 9.6 & 19 & 21.2 & 52.4 & 48.6\\
\cline{2-8}
& MRFR-C & 7.8 & 10.5 & 21 & 19.7 & 53 & 50.4\\
\cline{2-8}
& MRFR-TC & 6.5 & 9 & 17.5 & 19.5 & 50.7 & 46.5\\
\hline \hline
\multirow{4}{*}{2001} & MRFR & \textbf{10.2} & 8 & 20 & \textbf{22.8} & \textbf{55} & \textbf{60.3}\\
\cline{2-8}
& MRFR-T & 9.5 & 7.6 & 18 & 18.9 & 48.7 & 52.4\\
\cline{2-8}
& MRFR-C & 8.7 & \textbf{8.2} & \textbf{21} & 17.4 & 41 & 52.3\\
\cline{2-8}
& MRFR-TC & 8.5 & 7.5 & 16.3 & 17.4 & 44.6 & 50.8\\
\hline
\hline
\multirow{4}{*}{2002} & MRFR & \textbf{6.9} & \textbf{6.7} & \textbf{12.4} & \textbf{12.8} & \textbf{40.7} & \textbf{42.8}\\
\cline{2-8}
& MRFR-T & 6.4 & 6.2 & 10.7 & 11 & 37.4 & 37.9\\
\cline{2-8}
& MRFR-C & 6 & 5.7 & 10.3 & 9.4 & 36.4 & 37.6\\
\cline{2-8}
& MRFR-TC & 6.2 & 5.4 & 9.8 & 9 & 35.6 & 36\\
\hline
\hline
\multirow{4}{*}{2003} & MRFR & \textbf{5.4} & 5.2 & \textbf{10.2} & 9.7 & \textbf{24.6} & \textbf{28.6}\\
\cline{2-8}
& MRFR-T & 5.2 & \textbf{5.6} & 9.6 & 9.7 & 22.6 & 25.6\\
\cline{2-8}
& MRFR-C & 5 & 4.8 & 9.8 & \textbf{10.4} & 21.7 & 24\\
\cline{2-8}
& MRFR-TC & 4.8 & 4.8 & 9.6 & 10 & 20.5 & 24.4\\
\hline 
\end{tabular}
\end{table}

\textbf{Effect of parameters.} We conduct experiment to demonstrate how sensitive our model is to the parameters. Due to space limitation, we only report the results of the parameters associated with text features. For simplicity, let $\gamma_1=(1-\beta_p)(1-\alpha_p)$ denoting the paper-text feature graph parameter, and $\gamma_2=(1-\beta_a)(1-\alpha_a)$ denoting the author-text feature graph parameter. Fig. 6 shows the \emph{recommendation intensity} curves over various $\gamma_1$ and $\gamma_2$ when the other parameters are fixed. One can see that content information does improve the ranking performance. However, the performance decreases when $\gamma_1$ or $\gamma_2$ are too large, since the structure information is also important. The best performance is achieved at around $\gamma_1=0.4$ and $\gamma_2=0.3$.
\begin{figure}[!tp]
\centering
\setlength{\belowcaptionskip}{-0.5cm}
\setlength{\abovecaptionskip}{5pt}
  \includegraphics[height=2.7cm]{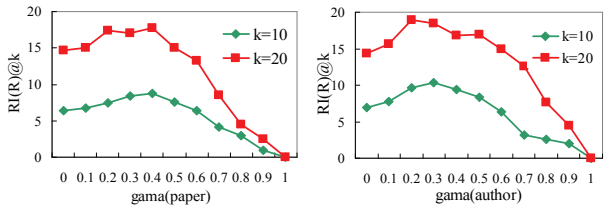}
  \caption{Effect of parameter $\gamma$}
\end{figure}
\section{Conclusion}
While most previous related works focused mainly on ranking the current importance of papers and authors, this paper proposed an approach MRFRank to predict the future influence of new publications and young researchers. MRFRank integrates the available time, graphs and rich texts information into a unified framework to rank papers and authors simultaneously. On the ArnetMiner dataset, we empirically evaluate our approach against state-of-the-art methods, and the results show the effectiveness of our approach. In the future, we will focus on: (1) how to improve the ranking performance by incorporating external resources such as social factors and best paper awards information; and (2) how to apply the model to other ranking scenarios like predicting potentially hot posts and influential users in social networks.

\section*{Acknowledgement}
This work is supported by the National Natural Science Foundation of China (Grant Nos. 61170189, 61370126, 61202239), the Research Fund for the Doctoral Program of Higher Education (Grant No. 20111102130003), the Fund of the State Key Laboratory of Software Development Environment (Grant No. SKLSDE-2013ZX-19), US NSF through grants CNS-1115234, DBI-0960443, and OISE-1129076, and US Department of Army through grant W911NF-12-1-0066.

\end{document}